\documentclass{elsart}
\usepackage[utf8]{inputenc}
\usepackage[T1]{fontenc}
\usepackage[english]{babel}
\usepackage{amssymb,amscd}
\usepackage{amsfonts}
\usepackage{amsmath}
\usepackage{xspace}
\usepackage{enumerate}
\usepackage{stmaryrd}
\usepackage{color}
\usepackage{graphicx}
\usepackage{tikz}
\usetikzlibrary{patterns}

\journal{Theoretical Computer Science}

\newcommand{\Z}{\mathbb Z}
\newcommand{\N}{\mathbb N}
\newcommand{\Ns}{\mathbb N\setminus\{0\}}

\newcommand{\deux}{\{0,1\}}
\newcommand{\sett}[2]{\left\{\left.#1\vphantom{#2}\right|#2\,\right\}}
\newcommand{\set}[3]{\sett{#1\in#2}{#3}}
\newcommand{\card}[1]{\left|#1\right|}

\newcommand{\abs}[1]{\left|#1\right|}

\newcommand{\cc}[2]{\left[#1,#2\right]}
\newcommand{\co}[2]{\left[#1,#2\right[}
\newcommand{\oc}[2]{\left]#1,#2\right]}
\newcommand{\scc}[2]{_{\cc{#1}{#2}}}
\newcommand{\rev}[1]{\widetilde{#1}}

\newcommand{\compl}[1]{\bar{#1}}

\newcommand{\soit}[1]{\left\{\everymath{\displaystyle\everymath{}}\begin{array}{ll}#1\end{array}\right.}
\newcommand{\si}{\textrm{ if }}
\newcommand{\sinon}{\textrm{ otherwise}}
\newcommand{\appl}[5]{\begin{array}{rcrcl}#1&:&#2&\to&#3\\&&#4&\mapsto&\displaystyle#5\end{array}}
\newcommand{\restr}[1]{_{\left|#1\right.}}

\newcommand{\spart}[1]{\left\lceil #1\right\rceil}
\newcommand{\ipart}[1]{\left\lfloor #1\right\rfloor}

\newcommand{\et}{\textrm{ and }}

\newcommand{\ie}{\emph{i.e.}\ }
\newcommand{\fig}[3]{\begin{figure}[htp]\centering\includegraphics[scale=#3]{#1.pdf}\caption{#2.}\end{figure}}

\begin{document}
\begin{frontmatter}

\title{Traced communication complexity\\ of cellular automata\thanksref{ecos}}
\thanks[ecos]{This work has been supported by the ECOS-Sud Project, Academy of Finland project 131558 (P. G.), Fondecyt 1070022, 1090156, BASAL-CMM (I. R.), Fondecyt 1100003, BASAL-CMM, Anillo Act88, SFI-Santa Fe (E. G.).}
\author[UAI]{Eric Goles}
\ead{eric.chacc@uai.cl}
\author[CMM,FUNDIM]{Pierre Guillon\corauthref{cor}}
\corauth[cor]{Corresponding author.}
\ead{pguillon@dim.uchile.cl}
\author[CMM]{Ivan Rapaport}
\ead{rapaport@dim.uchile.cl}
\address[UAI]{Facultad de Ingenier\'{\i}a y Ciencias
Universidad Adolfo Iba\~nez, Santiago, Chile}
\address[CMM]{DIM - CMM, UMI CNRS 2807, Universidad de Chile,
%Av. Blanco Encalada 2120, 8370459
Santiago, Chile}
\address[FUNDIM]{Department of Mathematics - University of Turku, Finland}

\begin{abstract}
We study cellular automata with respect to a new communication complexity problem: each of two 
players know half of some finite word, and must be able to tell whether the state of the central cell will follow a given evolution, by communicating as little as possible between each other. We present some links with classical dynamical concepts, especially equicontinuity, expansiveness, entropy and give the asymptotic communication complexity of most elementary cellular automata.
\end{abstract}


\end{frontmatter}

\section*{Introduction}

Cellular automata (CA) were introduced in the fifties in order to represent natural complex systems.
They were soon studied as a computational model, especially for parallelism. Indeed, they can be seen as wide networks of small machines communicating locally.

Introduced in \cite{yao}, the communication complexity of a function $f$ measures 
how much data must be exchanged between two machines which have part of the input of $f$ in order for one to be able to compute $f$.
This approach, representing the degree of parallelism needed, was adapted to the context of CA in \cite{comcplx1r} and gave interesting results in \cite{cc218,multiround,ccuniv} that help understand the computation represented by some CA. The complexity notion that these references use is somehow orthogonal to classical ones. For instance, bipermutive CA, which show strongly chaotic behaviors for many reasonable definitions, appear to be simple with respect to this complexity measure.

In this article, we study a new variant of communication complexity problem, which involves more links with dynamical chaos. In the first section, we give the definition of cellular automata and communication complexity. Then we address some properties of cellular automata which allow some simple protocols. The third and forth section are devoted to the links with trace, entropy and expansiveness. We finally discuss links with simulations.

\section{Definitions}
%In the whole article, we will fix a finite alphabet $A$.
%For a finite word $w\in A^*$, we note $\length w$ its length.
%A \emph{configuration} is a biinfinite word $x\in A^\Z$. We denote $x\scc ij$ (resp. $x\sco ij$) the finite word $x_i\ldots x_k$ (resp. $x_i\ldots x_{k-1}$).

\subsection{Cellular automata}
A CA consists in a sequence of cells with states in some alphabet $A$, evolving according to their neighbors.
We restrict our study to one-dimensional CA with nearest neighbors. This choice will be crucial for our definitions, but they still somehow apply to all one-dimensional CA since it is known that they can be simulated in a direct way by CA with nearest neighbors.
In this context, a \emph{cellular automaton} (CA) is a map $f:A^3\to A$. %; it can also be assimilated with the global map $\appl F\az\az x{(f(x_{i-1}x_ix_{i+1}))_{i\in\Z}}$.
In particular, if $A=\deux$, there are exactly 256 so-called \emph{elementary} CA, which can be referred to by the following canonical number: $\sum_{a,b,c\in\deux}f(abc)2^{4a+2b+c}$.

We denote $\cc ij$ the integer interval $\set k\Z{i\le k\le j}$.
In order to simplify notation, a \emph{configuration} $w$ is a finite word of odd length $2n+1$ whose indexes are centered around the origin, \ie $w=w_{-n}w_{-n+1}\ldots w_n\in A^{\cc{-n}n}$.
Moreover, if $\cc ij$ is a subinterval of an interval $I$ and $w\in A^I$ then $w\scc ij$ represents the pattern $w_i\ldots w_j$.
If $u\in A^{\cc ij}$ and $v\in A^{\oc jk}$ for some intervals $\cc ij,\oc jk\subset\Z$, then we will note $uv\in A^{\cc ik}$ the corresponding obvious juxtaposition of the two words.

The local rule of a CA can be applied in a parallel and synchronous way: $f:A^3\to A$ is extended to all $w\in A^{\cc ij}$ with $j-i\ge2$ by defining $\tilde f(w)\in A^{\cc{i+1}{j-1}}$ as $\tilde f(w)_k=f(w_{k-1},w_k,w_{k+1})$ for all $k\in\cc{i+1}{j-1}$; for convenience $\tilde f$ is again written as $f$.
We can therefore consider the iteration $f^t$ over any word $w\in A^{\cc ij}$ with $j-i\ge2t$.

Given an initial configuration $w\in A^{\cc{-n}n}$, its trace over a non-empty interval $\cc ij\subset\cc{-n}n$ of cells is the sequence of words $f^t(w)\scc ij\in A^{\cc ij}$ that one observes in the space-time diagram for input $w$. One has to restrict this to time steps $0\le t\le n-\max(\abs i,\abs j)$ for which the states of all cells in $\cc ij$ are defined.
With the exception of Section \ref{s:posexp}, in the paper only the special case $i=j$ is considered. One then gets:
\[\appl{T_f^{\{i\}}}{A^{\cc{-n}n}}{A^{\cc0{n-\abs i}}}w{(t\mapsto f^t(w)_i)~.}\]
We will write $T_f=T_f^{\{0\}}$, corresponding to the central column of the computation triangle.
It is a characteristic symbolic system linked to the CA dynamics, which was for instance proved very complex in \cite{utrace}.

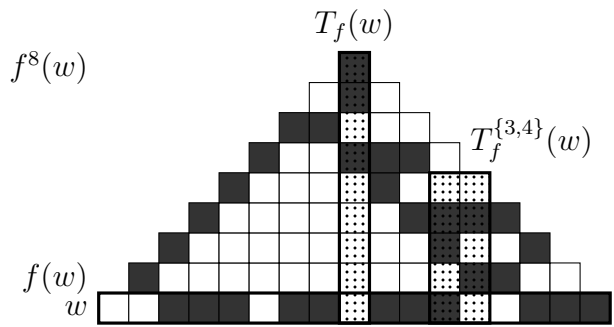
\begin{figure}[htp]\centering\begin{tikzpicture}[scale=.4]
\def\colb{black!80}\def\col{white}\def\contour{black}
\def\alice{}\def\bob{}\def\alices{}\def\bobs{}\def\gauche{}\def\droite{}\def\trace{}\def\centre{}\def\result{}
\def\ansc{gray}\def\alicec{blue}\def\bobc{orange}
 \tikzstyle 0=[fill=white]
 \tikzstyle 1=[fill=black!80]
 \newcommand{\cell}[3]{\draw[#3](#1,#2)rectangle+(1,1);}

\cell{0}{0}{0}\cell{1}{0}{0}\cell{2}{0}{1}\cell{3}{0}{1}\cell{4}{0}{1}\cell{5}{0}{0}\cell{6}{0}{1}\cell{7}{0}{1}\cell{8}{0}{0}\cell{9}{0}{1}\cell{10}{0}{1}\cell{11}{0}{1}\cell{12}{0}{0}\cell{13}{0}{0}\cell{14}{0}{1}\cell{15}{0}{1}\cell{16}{0}{1}
\cell{1}{1}{1}\cell{2}{1}{0}\cell{3}{1}{0}\cell{4}{1}{0}\cell{5}{1}{0}\cell{6}{1}{0}\cell{7}{1}{0}\cell{8}{1}{0}\cell{9}{1}{0}\cell{10}{1}{0}\cell{11}{1}{0}\cell{12}{1}{1}\cell{13}{1}{1}\cell{14}{1}{0}\cell{15}{1}{0}
\cell{2}{2}{1}\cell{3}{2}{0}\cell{4}{2}{0}\cell{5}{2}{0}\cell{6}{2}{0}\cell{7}{2}{0}\cell{8}{2}{0}\cell{9}{2}{0}\cell{10}{2}{0}\cell{11}{2}{1}\cell{12}{2}{0}\cell{13}{2}{0}\cell{14}{2}{1}
\cell{3}{3}{1}\cell{4}{3}{0}\cell{5}{3}{0}\cell{6}{3}{0}\cell{7}{3}{0}\cell{8}{3}{0}\cell{9}{3}{0}\cell{10}{3}{1}\cell{11}{3}{1}\cell{12}{3}{1}\cell{13}{3}{1}
\cell{4}{4}{1}\cell{5}{4}{0}\cell{6}{4}{0}\cell{7}{4}{0}\cell{8}{4}{0}\cell{9}{4}{1}\cell{10}{4}{0}\cell{11}{4}{0}\cell{12}{4}{0}
\cell{5}{5}{1}\cell{6}{5}{0}\cell{7}{5}{0}\cell{8}{5}{1}\cell{9}{5}{1}\cell{10}{5}{1}\cell{11}{5}{0}
\cell{6}{6}{1}\cell{7}{6}{1}\cell{8}{6}{0}\cell{9}{6}{0}\cell{10}{6}{0}
\cell{7}{7}{0}\cell{8}{7}{1}\cell{9}{7}{0}
\cell{8}{8}{1}
\draw[very thick, pattern=dots](8,0)rectangle(9,9);
\node[anchor=south]at (8.5,9){$T_f(w)$};
\draw[very thick, pattern=dots](11,0)rectangle(13,5);
\node[anchor=south west]at (12,5){$T_f^{\{3,4\}}(w)$};
\draw[very thick](0,0)rectangle(17,1);
\node[anchor=east]at (0,.5){$w$};
\node[anchor=east]at (0,1.5){$f(w)$};
\node[anchor=east]at (0,8.5){$f^8(w)$};
\end{tikzpicture}
  \caption{Some traces of the configuration $w=00101100000101110$ by CA rule $28$.}
\label{f:trace}
\end{figure}

\subsection{Communication complexity}
Let $f$ be a map defined over some Cartesian product $X\times Y$ into $A$. If $(x,y) \in X\times Y$, consider that two people, Alice and Bob, are given $x$ and $y$ respectively, and must compute the value $f(x,y)$ by communicating as little as possible between each other. This gives two variants of \emph{communication complexity} (CC).
The \emph{multi-round CC} of $f$ is the cost of the best communication protocol between the two players allowing one of them to produce the result $f(x,y)$. In other words, assume Alice and Bob agree on some deterministic protocol depending only on function $f$ (a sequence of data exchange from one to another where each message may depend on the previous ones); we are interested in the maximum for all possible inputs $x$ and $y$, of the number of bits exchanged between them, eventually allowing one of them to compute the result. The multi-round CC is the minimum of these maxima for all possible protocols they could have agreed on.

%A protocol is \emph{leftsided} if only Alice sends information to Bob.
The \emph{left} (\emph{one-round}) \emph{CC} is the worst case number of bits that Alice needs to send in order to allow Bob to directly compute the result; it corresponds to a protocol where only Alice sends information. We can similarly define the \emph{right CC}.
Of course, the multi-round CC is at most equal to the minimum between the two one-round CCs.

A function $f:\deux^n\times\deux^n\to\deux$ can be represented by the square matrix $M$ of size $2^n$ such that $M_{i,j}=f((u_{n-1}\ldots u_0),(v_0\ldots v_{n-1}))$ if $i=\sum_{k<n}u_k2^k$ and $j=\sum_{k<n}v_k2^k$. Then it is known (see for instance \cite{comcplx}) that the left (resp., right) CC of $f$ can be seen as the logarithm of the number of distinct lines (resp., columns) in the matrix $M$. Moreover, the multi-round CC is conjectured in \cite{rankcc} to be polylogarithmic in the rank of $M$. We will see some examples of such matrices, which help to get an intuition of the asymptotic CC: the complexity of the matrix in terms of number of distinct columns or lines is very visual.

A subset $S\subset X\times Y$ is a \emph{fooling set} for $f$ if for every two distinct pairs $(x,y),(x',y')\in S$, we have $f(x,y)=f(x',y')$ and either $f(x',y)$ or $f(x,y')$ is distinct from $f(x,y)$.

\begin{prop}[\cite{comcplx}]\label{p:fooling}
If a function $f$ admits a fooling set $S$, then its multi-round CC is lower-bounded by  $\log{\card S}$.
\end{prop}

\subsection{Traced communication complexity}

We can actually view a CA $f$ as computing a function. The sequence of words $f^t(w)$ for $0\le t<n$ and $w=w_{-n}\ldots w_n\in A^{\cc{-n}n}$ (drawn in Figure \ref{f:trace}%) that we picturize centering the central cell of each word one above the other
) is the \emph{computation triangle}. If we consider that Alice initially knows $w_{-n}\ldots w_0$ and Bob $w_0\ldots w_n$, it is obvious that the former can compute the word $f^t(w)\scc{t-n}{-t}$ and the latter the word $f^t(w)\scc t{n-t}$, for $0\le t\le \ipart{\frac n2}$. The other parts of the computation triangle will a priori require information exchange between Alice and Bob.

In \cite{comcplx1r,cc218}, CC has been applied to CA, basically as that of the computation of the top cell, \ie the CC of the map $(w\scc{-n}{-1},w\scc1n)\mapsto f^n(w)$, for any $n\in\N$, where Alice and Bob share some fixed $w_0$.
We will refer to this notion as \emph{classical CC}.
In \cite{multiround}, this notion was generalized to arbitrary cutting position (not always $0$) between Alice's word and Bob's.
In \cite{ccuniv}, some new problems, the so-called \emph{invasion} and \emph{cycle length}, were associated to the CA.
In each of these, the idea is to consider the CA as complex (resp. simple) if the CC is asymptotically linear (resp. constant) when the size $n$ of the input grows to infinity.

In a binary alphabet, the classical problem is for Alice and Bob to know whether the top cell of the computation triangle is a $1$. Instead of this we can require them to determine whether some state $1$ appears in the central column, \ie whether $f^t(w)_0=1$ for some step $t\in\cc0n$.
More generally, having fixed an arbitrary alphabet $A$, an integer $n\in\N$ and a word $z\in A^{n+1}$, consider the indicator function:
\[\appl{\hat f_z}{A^{\cc{-n}1}\times A^{\cc1n}}\deux{(u,v)}{\soit{0&\si T_f(uz_0v)=z\\1&\sinon.}}\]
Its CC will be referred to as the \emph{traced CC}.

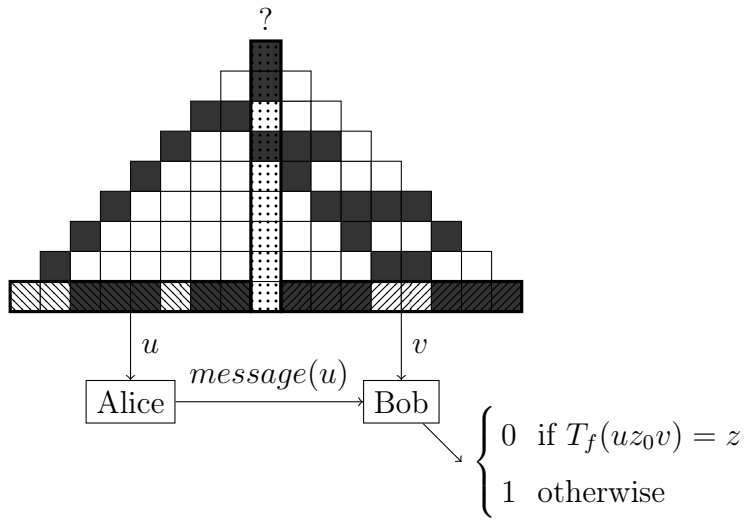
\begin{figure}[htp]\centering\begin{tikzpicture}[scale=.4]
\def\colb{black!80}\def\col{white}\def\contour{black}
\def\alice{}\def\bob{}\def\alices{}\def\bobs{}\def\gauche{}\def\droite{}\def\trace{}\def\centre{}\def\result{}
\def\ansc{gray}\def\alicec{blue}\def\bobc{orange}
 \tikzstyle 0=[fill=white]
 \tikzstyle 1=[fill=black!80]
 \newcommand{\cell}[3]{\draw[#3](#1,#2)rectangle+(1,1);}

\cell{0}{0}{0}\cell{1}{0}{0}\cell{2}{0}{1}\cell{3}{0}{1}\cell{4}{0}{1}\cell{5}{0}{0}\cell{6}{0}{1}\cell{7}{0}{1}\cell{8}{0}{0}\cell{9}{0}{1}\cell{10}{0}{1}\cell{11}{0}{1}\cell{12}{0}{0}\cell{13}{0}{0}\cell{14}{0}{1}\cell{15}{0}{1}\cell{16}{0}{1}
\cell{1}{1}{1}\cell{2}{1}{0}\cell{3}{1}{0}\cell{4}{1}{0}\cell{5}{1}{0}\cell{6}{1}{0}\cell{7}{1}{0}\cell{8}{1}{0}\cell{9}{1}{0}\cell{10}{1}{0}\cell{11}{1}{0}\cell{12}{1}{1}\cell{13}{1}{1}\cell{14}{1}{0}\cell{15}{1}{0}
\cell{2}{2}{1}\cell{3}{2}{0}\cell{4}{2}{0}\cell{5}{2}{0}\cell{6}{2}{0}\cell{7}{2}{0}\cell{8}{2}{0}\cell{9}{2}{0}\cell{10}{2}{0}\cell{11}{2}{1}\cell{12}{2}{0}\cell{13}{2}{0}\cell{14}{2}{1}
\cell{3}{3}{1}\cell{4}{3}{0}\cell{5}{3}{0}\cell{6}{3}{0}\cell{7}{3}{0}\cell{8}{3}{0}\cell{9}{3}{0}\cell{10}{3}{1}\cell{11}{3}{1}\cell{12}{3}{1}\cell{13}{3}{1}
\cell{4}{4}{1}\cell{5}{4}{0}\cell{6}{4}{0}\cell{7}{4}{0}\cell{8}{4}{0}\cell{9}{4}{1}\cell{10}{4}{0}\cell{11}{4}{0}\cell{12}{4}{0}
\cell{5}{5}{1}\cell{6}{5}{0}\cell{7}{5}{0}\cell{8}{5}{1}\cell{9}{5}{1}\cell{10}{5}{1}\cell{11}{5}{0}
\cell{6}{6}{1}\cell{7}{6}{1}\cell{8}{6}{0}\cell{9}{6}{0}\cell{10}{6}{0}
\cell{7}{7}{0}\cell{8}{7}{1}\cell{9}{7}{0}
\cell{8}{8}{1}
\draw[very thick, pattern=north west lines](0,0)rectangle(8,1);
\node[rectangle,draw] (alice) at (4,-3){Alice};
\draw[->](4,0) to node[right] {$u$} (alice);
\draw[very thick, pattern=dots](8,0)rectangle(9,9);
\node[anchor=south]at (8.5,9){?};
\draw[very thick, pattern=north east lines](9,0)rectangle(17,1);
\node[rectangle,draw] (bob) at (13,-3){Bob};
\draw[->](13,0) to node[right] {$v$} (bob);
\draw[->](alice) to node[above] {$message(u)$} (bob);
\node[anchor=west] (res) at (15,-5) {$\soit{0&\si T_f(uz_0v)=z\\1&\sinon}$};
\draw[<-](15,-5) to node{} (bob);
\end{tikzpicture}
  \caption{One-round protocol for the traced CC.}
\end{figure}
Note that the rule $f$ and the word $z$ are fixed, hence allowing Alice and Bob to agree on a protocol which will depend on them.
We will see some examples of simple reasoning on the local rule to bound the CC corresponding to some given word $z$, often the uniform word $z=0^{n+1}$.
We then study some properties inspired by topological dynamics which imply either low CC for any word $z$ or high CC for some word. The idea is more or less to consider a CA as (extremely) complex if there is a linear map of $n$ which lower-bounds the maximal multi-round CC corresponding to the words of length $n$. On the contrary, it will be considered as (extremely) simple if any such CC is bounded by a constant.
When considering the word $0^{n+1}$, note that (each version of) the CC $\hat f_{0^{n+1}}$ is nondecreasing on $n$, since if a $1$ appears in the trace within $n$ steps, then in particular it appears within $n+1$ steps. We will abusively say that the CC of $\hat f_{0^{n+1}}$ is constant when it is asymptotically, \ie when it is bounded by some constant.

% we study some classes of CA for which we can say whether this CC is asymptotically low or high in terms of $n$.
% We will sometimes emphasize the uniform case $\hat f_{0^{n+1}}$.

In the next sections, an alphabet $A$, an integer $n\in\N$ and a word $z\in A^{n+1}$ are fixed, unless explicitly stated otherwise. Each figure presented further will represent, for some elementary CA $f$, the matrix of the function $\hat f_{0^{n+1}}$ (assimilating gray with black), superposed with the corresponding matrix corresponding to the classical CC (assimilating white with gray); in other words, gray cells correspond to the words whose evolution has reached state $1$ but eventually came back to state $0$.

\section{Simple communications}
\subsection{One-sided rules}
Similarly to the classical CC, the traced CC of any one-sided CA, \ie those that depend only on either left cells or right cells, is clearly constant: one of the two parties is completely able to compute the function $\hat f_z$ by himself.
We can be a little more general in the (nearly) uniform case.

If $B\subset A$, then a CA $f$ is \emph{$B$-leftsided} if $\forall a,c,d\in A,b\in B,f(abc)=f(abd)$.
Similarly, we define \emph{$B$-rightsided} CA. A CA is \emph{$B$-onesided} if it is either $B$-leftsided or $B$-rightsided.
If all letters of $z$ are in $B$, one party can compute the evolution of the central cell if it stays in $B$; if it does not, then the trace cannot be $z$. Hence, no communication is needed.
\begin{prop}\label{p:onesided}
 For any $B$-onesided CA $f$, the one-round CC for $\hat f_z$, with $z\in B^{n+1}$, is at most $1$.
\end{prop}
\begin{pf*}{Proof.}
Let us prove the result for left CC; the case of right CC is symmetric.
\begin{itemize}
 \item If it is $B$-leftsided, Alice can compute whether some letter which is not in $B$ appears in the central cell and give the answer to Bob.
 \item If it is $B$-rightsided, she does not say anything to Bob; he will be able to find the answer by himself.
\qed\end{itemize}
\end{pf*}
 For $z=0^{n+1}$, the previous proposition can be applied to the 64 $0$-onesided elementary CA ($B$-onesided for $B=\{0\}$), \ie those whose number can be written as $a_7a_6a_4a_4a_3a_2a_0a_0$ or $a_7a_6a_1a_0a_3a_2a_1a_0$ in base $2$. %µ 64=8, 68=12, 192=136, 196=140
\fig{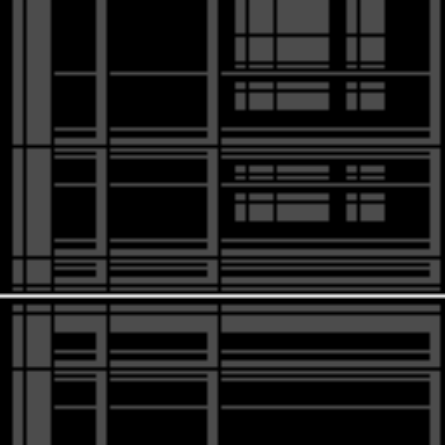}{Matrix of the rule 143}{.5}

\subsection{Spreading states}
A state $0\in A$ is \emph{quiescent} for the CA $f$ if $f(000)=0$.
Consider a subalphabet $B\subset A$. %is \emph{quiescent} for CA $f$ if $f(B^3)\subset B$. In that case the restriction $f\restr B$ to this subalphabet acts as a \emph{subautomaton}.
We note $\compl B$ the complementary subalphabet $A\setminus B$ (we may also note $\compl0=A\setminus\{0\}$).
We say that $B$ is \emph{left semi-strongly spreading} (resp., \emph{weakly spreading}) for the CA $f$ if $f(A\compl BB)\subset B$ (resp., $f(\compl B\compl BB)\subset B$). Symmetrically, right spreadingness can be defined.

We can see that any elementary CA for which state $1$ is both left and right semi-strongly spreading and state $0$ is quiescent has $\hat f_{0^{n+1}}(u,v)=1$ if and only if $u0v$ contains some $1$, which will progressively spread towards the center. The following proposition generalizes this observation.
% \begin{prop}
%  If $B\subset A$ is left and right semi-strongly spreading, $\compl B$ quiescent and $z\in\compl B^*$, then the right CC for $\hat f_z$ is equal - up to a constant - to that of $\hat f_z$ over the subautomaton $f\restr{\compl B}$.
% \end{prop}
% \begin{pf*}{Proof.}
% If Bob has a letter in $B$, then he can say with one bit to Alice that $\hat f$ will give $1$; otherwise he uses the protocol for the restriction $f\restr{\compl B}$; Alice will either have a letter in $B$ or use this protocol to find the answer.
% \qed\end{pf*}
\begin{prop}
 If $0$ is quiescent and $\compl0$ is left semi-strongly spreading, then the right CC for $\hat f_{0^{n+1}}$ is at most $1$.
\end{prop}
\begin{pf*}{Proof.}
If Bob has a letter in $\compl0$, then he knows that it will spread towards the center, and he can say with one bit to Alice that $\hat f$ will give $1$. Otherwise Alice knows that he has word $v=0^n$.
\qed\end{pf*}
Symmetrically, if $0$ is quiescent and $\compl0$ is right semi-strongly spreading, then the left CC for $\hat f_{0^{n+1}}$ is constant.
On the other hand, if $f$ is an elementary CA such that $\compl0$ is semi-strongly spreading but $0$ is not quiescent, then a rapid case study shows that we always have $\hat f_{0^{n+1}}=1$ as soon as $n>1$. %cas simple pour la 147 (2 blancs à la suite) / 167 (3)
We globally obtain, whenever $\compl0$ is semi-strongly spreading, a constant CC for $\hat f_{0^{n+1}}$. This corresponds to the 96 elementary CA whose number in base $2$ can be written either as $a_7a_611a_3a_2a_1a_0$ or as $a_7a_61a_4a_3a_21a_0$.
\fig{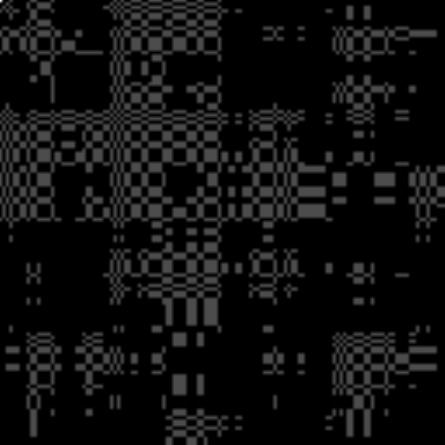}{Matrix of the rule 182}{.5}

\subsection{Stagnating states}
 A word $u\in A^*$ is \emph{stagnating} for the CA $f$ if $\forall a,b\in A,f(aub)=u$.

In this subsection, we assume $A=\deux$.
Note that if $0$ is stagnating, then the rule is both $0$-leftsided and $0$-rightsided and we have already seen that the one-round CC is constant. We can generalize this to the following case.
\begin{prop}
 If $0$ is quiescent and $1$ is neither left nor right weakly spreading, then both one-round CCs for $\hat f_{0^{n+1}}$ are at most $1$.
\end{prop}
\begin{pf*}{Proof.}\begin{itemize}
\item If $f(101)=0$, then $0$ is stagnating; this is a subcase of Proposition \ref{p:onesided}.
\item If $f(101)=1$, then $00$ is stagnating, but single $0$s disappear. Hence $\hat f(u,v)=0$ if and only if $u_{-1}=0$ or $v_1=0$. The result of this test can be transmitted in one bit.
\qed\end{itemize}\end{pf*}
This proposition applies to the elementary CA who map $0$ to any neighborhood containing two consecutive $0$s, \ie whose number in base $2$ can be written $a_7a_6a_50a_3a_200$. %: 32, 36, 40, 44, 104, 108, 160, 164, 168, 172, 232, 236. +96, 100
\fig{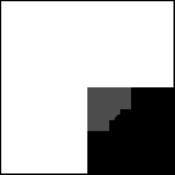}{Matrix of the rule 232}{1.2}

\begin{rem}\label{r:stagn1}
 Any elementary CA $f$ with stagnating $1$ has a CC for $\hat f_{0^{n+1}}$ equal to its classical CC.
Indeed, for any words $u\in A^{\cc{-n}{-1}}$ and $v\in A^{\cc1n}$, $\hat f_{0^{n+1}}(u,v)$ is equal to $f^n(u0v)$.
For instance, from \cite{ccuniv}, the CA 222 has logarithmic CC for $\hat f_{0^{n+1}}$.
 \end{rem}
\fig{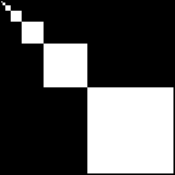}{Matrix of the rule 222}{1.2}

\section{Trace protocol}
In order to decide whether $T_f(uz_0v)=z$ or not, Alice and Bob need to compute only until the first step $t\in\cc1n$ when the central state does not correspond to $z_t$. In a first approximation, each of them can compute their side by assuming that the central column corresponds to $z$ -- note that this central column and his/her initial word completely determine his/her half-triangle, since the rule has radius $1$ -- and then check if this assumption could lead to a contradiction.

Formally, for $v\in A^{\cc1n}$ we define $f_{\to z}^0(v)=v$ and for $t<n-1$, $f_{\to z}^{t+1}(v)=f(z_tf_{\to z}^t(v))$, which is still an element of $A^{\cc1n}$. %(with indexes from $1$ to $n$).
Then $T_{f\to z}(v)=(f_{\to z}^t(v)_1)_{t<n}$ represents the column that should be just right to the column $z$ in a computation triangle where Bob has word $v$ (if ever such a triangle is possible).

\begin{figure}[htp]\centering\begin{tikzpicture}[scale=.4]
\def\colb{black!80}\def\col{white}\def\contour{black}
\def\alice#1{}\def\bob{}\def\alices#1{}\def\bobs{}\def\gauche#1{}\def\droite{}\def\trace{}\def\centre{}\def\result{}
\def\ansc{gray}\def\alicec{blue}\def\bobc{orange}
 \tikzstyle 0=[fill=white]
 \tikzstyle 1=[fill=black!80]
 \newcommand{\cell}[3]{\draw[#3](#1,#2)rectangle+(1,1);}

\cell{0}{0}{0}\cell{1}{0}{0}\cell{2}{0}{1}\cell{3}{0}{1}\cell{4}{0}{1}\cell{5}{0}{0}\cell{6}{0}{1}\cell{7}{0}{1}\cell{8}{0}{0}\cell{9}{0}{1}\cell{10}{0}{1}\cell{11}{0}{1}\cell{12}{0}{0}\cell{13}{0}{0}\cell{14}{0}{1}\cell{15}{0}{1}\cell{16}{0}{1}
\cell{1}{1}{1}\cell{2}{1}{0}\cell{3}{1}{0}\cell{4}{1}{0}\cell{5}{1}{0}\cell{6}{1}{0}\cell{7}{1}{0}\cell{8}{1}{0}\cell{9}{1}{0}\cell{10}{1}{0}\cell{11}{1}{0}\cell{12}{1}{1}\cell{13}{1}{1}\cell{14}{1}{0}\cell{15}{1}{0}
\cell{2}{2}{1}\cell{3}{2}{0}\cell{4}{2}{0}\cell{5}{2}{0}\cell{6}{2}{0}\cell{7}{2}{0}\cell{8}{2}{0}\cell{9}{2}{0}\cell{10}{2}{0}\cell{11}{2}{1}\cell{12}{2}{0}\cell{13}{2}{0}\cell{14}{2}{1}
\cell{3}{3}{1}\cell{4}{3}{0}\cell{5}{3}{0}\cell{6}{3}{0}\cell{7}{3}{0}\cell{8}{3}{0}\cell{9}{3}{0}\cell{10}{3}{1}\cell{11}{3}{1}\cell{12}{3}{1}\cell{13}{3}{1}
\cell{4}{4}{1}\cell{5}{4}{0}\cell{6}{4}{0}\cell{7}{4}{0}\cell{8}{4}{0}\cell{9}{4}{1}\cell{10}{4}{0}\cell{11}{4}{0}\cell{12}{4}{0}
\cell{5}{5}{1}\cell{6}{5}{0}\cell{7}{5}{0}\cell{8}{5}{1}\cell{9}{5}{1}\cell{10}{5}{1}\cell{11}{5}{0}
\cell{6}{6}{1}\cell{7}{6}{1}\cell{8}{6}{0}\cell{9}{6}{0}\cell{10}{6}{0}
\cell{7}{7}{0}\cell{8}{7}{1}\cell{9}{7}{0}
\cell{8}{8}{1}
\draw[very thick, pattern=dots](8,0)rectangle(9,9);
\node[anchor=south]at (8.5,9){$z$};
\draw[very thick, pattern=north east lines](9,0)rectangle(17,1);
\node[anchor=west]at (17,.5){$v$};
\draw[very thick, pattern=crosshatch](9,0)rectangle(10,8);
\node[anchor=south west]at(9.5,8){$T_{f\to z}(v)$};
\end{tikzpicture}
  \caption{Trace protocol in Bob's side.}
\end{figure}
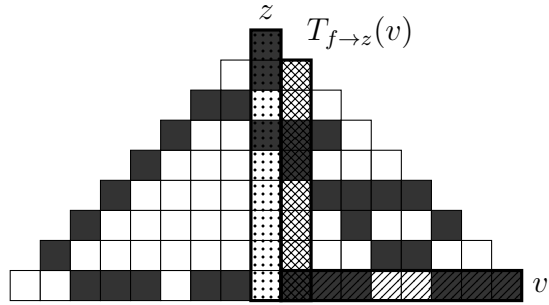

The word $T_{f\to z}(v)$ actually represents a valid message from Bob to Alice. In particular, if it has a short algorithmic complexity, then the right CC will be low.
\begin{prop}\label{p:otherbound}
 The right CC for $\hat f_z$ is upper-bounded by $\spart{\log\card{T_{f\to z}(A^{\cc1n})}}$.
\end{prop}
\begin{pf*}{Proof.}
If Alice has word $u$, Bob word $v$ and $m$ is such that for any step $t<m$, the central letter of $f^t(uz_0v)$ is $z_t$, then by an immediate recurrence, we have $\forall t<m,f^{t+1}(uz_0v)_0=f(T_{f\leftarrow z}(u)_tz_tT_{f\to z}(v)_t)$.
In $\spart{\log\card{T_{f\to z}(A^{\cc1n})}}$ bits, Bob can encode the data of the word $T_{f\to z}(v)$, and give it to Alice.
She can then compute $T_{f\leftarrow z}(v)$; she answers that $\hat f_z(u,v)=0$ if and only if the successive central cells represent a valid application of the local rule when juxtaposing the three columns.
\end{pf*}

Let us see a little variant of the previous result. Consider the set $\tau_{f\to z}=\sett{T^{\{1\}}_f(w)}{w\in A^{\cc{-n}n}\et T_f(w)=z}$ of all possible columns juxtaposed at the right of $z$ in some (full valid) computation triangle. Note that all of its elements can be written as $T_{f\to z}(w\scc1n)$, but the converse is false in the sense that the right half-triangle built with some $T_{f\to z}(v)$ could be impossible to extend to the left.
\begin{prop}\label{p:sentr}
 The right CC for $\hat f_z$ is upper-bounded by $\spart{\log\left(\card{\tau_{f\to z}}+1\right)}$.
\end{prop}
\begin{pf*}{Proof.}
Looking back at the previous proof, note that Bob can encode in $\spart{\log\left(\card{\tau_{f\to z}}+1\right)}$ bits the data of the word $T_{f\to z}(v)$ in the case when it belongs to $\tau_{f\to z}$, and some prespecified extra code otherwise.
In the first case, Alice will be able to compute $\hat f_z(u,v)$ as before.
In the second case, Alice will know that there is no possible initial word of right part $v$ which can give $z$ as a trace, hence $\hat f_z(u,v)=0$.
\qed\end{pf*}
Symmetrically, we can define $T_{f\leftarrow z}$ and $\tau_{f\leftarrow z}$ in Alice's side.
The left CC for $\hat f_z$ is upper-bounded by $\spart{\log\card{T_{f\leftarrow z}(A^{\cc1n})}}$, and by $\spart{\log\left(\card{\tau_{f\leftarrow z}}+1\right)}$.

\subsection{Grouping}
\newcommand{\bk}[1]{^{<#1>}}
Let us see a simple class of CA where Alice and Bob's sides can be seen independently enough, each one with the information of the trace, to get a very low CC: they do not even need to send their $T_{f\to z}$ or $T_{f\leftarrow z}$ words to one another.

Let $f:A^3\to A$ be a CA. We define its $2$-grouped as the CA $f\bk2$ on alphabet $A^2$ defined by $f\bk2((x_{-1},y_{-1}),(x_0,y_0),(x_1,y_1))=(f(y_{-1},x_0,y_0),f(x_0,y_0,x_1))$. It can be seen as the same CA where the cells have been grouped $2$ by $2$.

Grouping is one of the interesting tools allowing to define cellular simulation and intrinsic universality, which are a way to order the CA in terms of their ability to embed the dynamics of other CA (see for instance \cite{bulk2}). The classical CC of CA has been used to prove the non-ability of a CA to simulate other CA, based on the fact that this CC is nonincreasing with simulation. We will see that it is no more the case for traced CC.
\begin{prop}\label{p:bulk}
For any $2$-grouped CA $f=g\bk2$ and any word $z\in A^{n+1}$, the one-round CC of $\hat f_z$ is at most $1$.
\end{prop}
\begin{pf*}{Proof.}
It can be seen that $T_f(uz_0v)=z$ if and only if both $T_{f\leftarrow z}(u)\in\tau_{f\leftarrow z}$ and $T_{f\to z}(v)\in\tau_{f\to z}$: the $2$-block construction allows to glue any two valid half-triangles together into a full valid computation triangle.
If Alice has word $u$ and Bob $v$, he can compute $T_{f\to z}(v)$ and check whether it belongs to $\tau_{f\to z}$. It only needs to send the result of this test to Alice.
\qed\end{pf*}
As a result, even the simplest simulation, \ie the reverse operation of $2$-grouping, can increase the traced CC.

\subsection{$B\star$onesided rules}
Let us see a simple case where Proposition \ref{p:sentr} can be applied, which corresponds to some other kind of partial onesidedness.
If $B\subset A$, then a CA $f$ is \emph{$B\star$leftsided} if $\forall b,c,d\in A,\forall a\in B,f(abc)=f(abd)$.
Similarly, we define \emph{$B\star$rightsided} CA. %A CA is \emph{$0$-onesided} if it is either $0$-leftsided or $0$-rightsided.
\begin{prop}
 If $z\in B^{n+1}$ and $f$ is $B\star$leftsided, then the right CC for $\hat f_z$ is constant.
\end{prop}
\begin{pf*}{Proof.}
For any two words $w,w'\in A^{\cc{-n}n}$, if $T_f(w)=T_f(w')=z$ and $w_1=w'_1$, then by the $B\star$leftsided property we have $f(w)_1=f(w')_1$, and by induction, for any $t<n$, $f^t(w)_1=f^t(w')_1$.
It results that the words of $T_{f\to z}(A^{\cc1n})$ are determined by their first letter, which gives $\card{T_{f\to z}(A^{\cc1n})}=\card A$.
From Proposition \ref{p:otherbound}, the right CC for $\hat f_z$ is then bounded by $\spart{\log\card A}$.
\qed\end{pf*}
Intuitively, if Bob has the word $v$, then it is sufficient for him to send to Alice the initial state $v_1$ of his first cell, since the evolution of this cell will not depend on cells which are on the right if he assumes that the central column is $z$. Alice can thus compute $T_{f\to z}(v)$ and then know whether the central cell will reach some state which does not correspond to $z$.

Similarly, if $f$ is $B\star$rightsided, then the left CC for $\hat f_z$ is constant.
Overall, the proposition applies to the elementary CA whose number in base $2$ can be written $a_7a_6a_5a_4a_2a_2a_0a_0$ or $a_7a_2a_5a_0a_3a_2a_1a_0$.
% The following elementary CA (and their symmetric variants) are concerned by one of these conditions for $B=\{0\}$ ($z=0^{n+1}$): 19, 27, 31, 95, 147, 155, 159, 223. %27=83, 31=87, 155=211, 159=215
\fig{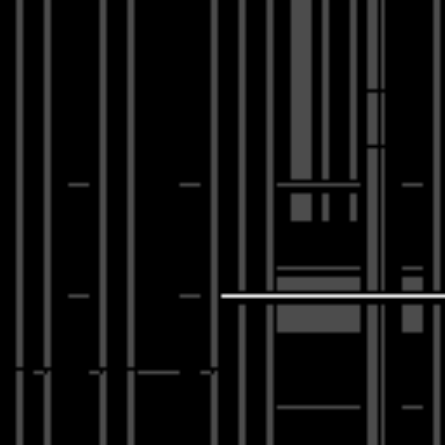}{Matrix of the rule 159}{.5}

The hypothesis is satisfied in two cases: the cell just on the right of the word $z$ has a state which either loops, or remains always the same. In the latter case and for a binary alphabet, we can be slightly more general if we allow logarithmic communications.
\begin{prop}%µ también elementary
 If $\exists a\in\deux,f(0a0)=f(0a1)=a$, then the right CC for $\hat f_{0^{n+1}}$ is upper-bounded by $\spart{\log(n+1)}$.
\end{prop}
\begin{pf*}{Proof.}
Note that $\appl\gamma{T_{f\to0^{n+1}}(A^{\cc1n})}{\co0n\cup\{+\infty\}}{z'}{\min_{z'_t=a}t}$ is injective, with inverse $\gamma^{-1}(t)=\compl a^ta^{n-t}$ if $t<n$, $\compl a^n$ if $t=+\infty$. Hence $T_{f\to0^{n+1}}(A^{\cc1n})\le n+1$, and we conclude thanks to Proposition \ref{p:otherbound}.
\qed\end{pf*}
Intuitively, if Bob has the word $v$, then it is sufficient for him to send to Alice the first generation when some $a$ appears in $T_{f\to0^{n+1}}(v)$ and $+\infty$ if $a$ never appears. Indeed, Alice will then know entirely this word, and be able to compute the result.

Similarly, if $\exists a\in\deux,f(0a0)=f(1a0)=a$, then the left CC for $\hat f_{0^{n+1}}$ is at most logarithmic.
Note that the previous case includes that, already seen, of CA having stagnating $0$, or stagnating $1$, and more generally any elementary CA whose number in base $2$ can be written $a_7a_6a_5a_4a_3a_2a_0a_0$ or $a_7a_6a_5a_4a_2a_2a_1a_0$. %as well as the following other CA: 41, 45, 105, 109, 169, 173, 233, 237. %41=97, 45=101, 169=225, 173=229
% 45, 173, 109 et 237 paraissent constantes
\fig{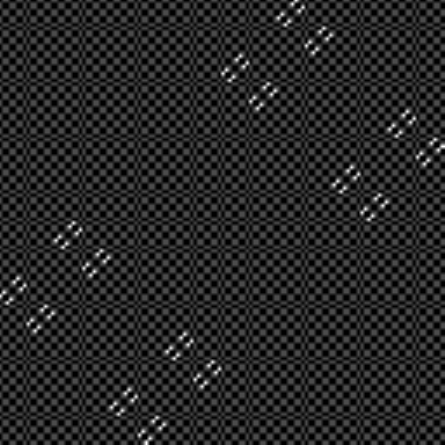}{Matrix of the rule 105}{.5}

\subsection{Entropy}
The next notions come from topological dynamics, but we emphasize here the point of view based on the trace of configurations as finite words.
Even though it is not crucial, the logarithms will be assumed binary.

The \emph{entropy} of the trace over cells $\co ij\subset\Z$ is the limit \[h_{j-i}=\lim_{n\to\infty}\frac{\log\card{T_f^{\co ij}(A^{\cc{-n}n})}}{n%+1-\max(\abs i,\abs j)
}~.\]
Thanks to the parallelism of the rule application, it only depends on the difference $j-i$ (and on the CA).
The \emph{entropy} of the CA $f$ is the supremum $h$ of the entropies of the traces over $\co ij$, when $j-i$ grows. This notion represents somehow the degree of ``disorder'' in the apparent evolution of the CA. %It is known to be upper-bounded by $\card A^2$.

Proposition \ref{p:sentr} gives us the following rough upper bound on the CC of $\hat f_z$: $\spart{\log\left(\card{T^{\{1\}}_f(A^{\cc{-n}n})}+1\right)}$. %Indeed, if $T^{\{1\}}_f(A^{\cc{-n}n})=T_{f\to z}(A^n)$, then the additional bit is trivial, so we can get rid of it in the protocol; otherwise $T_f(A^{\cc{-n}n})\supsetneq T_{f\to z}(A^n)$.
This allows us to state that null-entropy CA have sublinear one-round CC for $\hat f_z$. On the contrary, for CA of entropy $h>0$, it is known that the entropy $h_{j-i}$ of any nontrivial trace over $\co ij$ is also strictly positive, and in that case the one-round CC for $\hat f_z$ and for a large $n$ is at most $h_1n+o(1)$ which itself is at most $hn+o(1)$.

The inequality $h_{j-i}n\le hn$ corresponds to an interesting open problem on the structure of CA computations: whether there exists a computable bound on $\cc ij$ (maybe width $2$) such that $h_{j-i}=h$ (see for instance \cite{rowss}). The inequality $c\le h_{j-i}n+o(1)$ is more specific to our problem. It is not tight at all because the entropy gives intuition about the disorder visible in a finite window of the computation, without distinguishing whether the disorder comes from one single side or both, which is our purpose. For instance, onesided CA could give rise to a complex trace -- like the shift CA, for which $T_f(A^{\cc{-n}n})$ is the whole $A^{n+1}$ -- but we have already seen that their CC is simple.

\subsection{Equicontinuity}
A CA $f$ is \emph{equicontinuous} if for any $\cc ij\subset\Z$, the cardinality of $T_f^{\cc ij}(A^{\cc{-n}n})$, with $n\ge\max(\abs i,\abs j)$, is bounded by a constant. From \cite{KurkCA}, this notion corresponds to ultimately periodic CA, or equivalently to those for which the width-$1$ trace $T_f(A^{\cc{-n}n})$ has bounded cardinality for $n\in\N$.
This notion represents an extreme stability of the system, since distant cells cannot influence each other. Such CA were proved to have simple classical CC in \cite{ccuniv}. Here too, thanks to Proposition \ref{p:sentr}, we can see that equicontinuous CA have a constant one-round CC for $\hat f_z$.

\section{Expansiveness}\label{s:posexp}
We now deal with CA presenting some kind of complexity, which will give us high CC.
At the extreme opposite of equicontinuity, we say that a CA $f$ is (\emph{positively}) \emph{right-expansive} if there exists some time step $t^\to_f\in\Ns$ such that any two words % (``vertical'') word $z\in T_f^{\cc{-1}0}(A^{\cc{-t^\to_f}{t^\to_f}})$, there exists a unique $a\in A$ such that any $w%=w_{\cc{-t^\to_f}{t^\to_f}}
$w,w'\in A^{\cc{-t^\to_f}{t^\to_f}}$ with the same trace $T^{\cc{-1}0}_f(w)=T^{\cc{-1}0}_f(w')$ have the same letter $w_1=w'_1$ just on the right. In other words, being given the trace, we can rebuild -- in a unique way -- the right part of the initial finite word.

The simplest right-expansive CA are the \emph{right-permutive} ones, \ie the rules $f:A^3\to A$ such that $\forall a,b,c,d\in A,c\ne d\Rightarrow f(abc)\ne f(abd)$. It can be noted that they correspond to $t^\to_f=1$. In particular, for any $u\in A^{\cc{-n}{-1}}$, the restriction of $T_f$ over the set $uA^{\cc0n}$ is a bijection onto $A^{n+1}$. A well-known example is rule 90, which acts as an exclusive-or gate over the two extreme neighbors.

The notion of expensiveness is rather precise, but we can generalize it to some kind of subsystems of CA, in order for our lower bounds of CC to concern more CA, since it is intuitive that a system is at least as complex as its subsystems. Let us then define, in our setting, what corresponds to the symbolic notion of subshift of finite type.
If $\mathcal F\subset A^*$ is a finite language of \emph{forbidden patterns} and $\cc ij\subset\Z$, we note \[\Sigma\scc ij=\Sigma^{\mathcal F}\scc ij=\set{w_i\ldots w_j}{A^{\cc ij}}{\forall\cc{i'}{j'}\subset\cc ij,w\scc{i'}{j'}\notin\mathcal F}~.\]
%and $\Sigma=\Sigma^{\mathcal F}=\bigcup_{k\in\N}\Sigma^{\mathcal F}_k$ the set of words avoiding all patterns of $\mathcal F$.
%Such sets correspond to a \emph{subshift of finite type} in symbolic dynamics.
Consider the restriction $f\restr\Sigma$ of the extended rule of the CA $f$ to $\bigcup_{\cc ij\subset\Z}\Sigma\scc ij$. It is called a \emph{subautomaton} if this set is stable, \ie $f$ does not create forbidden patterns: $\forall\cc ij\subset\Z,f(\Sigma^{\mathcal F}\scc{i-1}{j+1})\subseteq\Sigma^{\mathcal F}\scc ij$.

%We can generalize the notion of expensiveness to all the subautomata of CA:
The subautomaton $f\restr\Sigma$ is \emph{right-expansive} if there exists some time $t^\to_f\in\Ns$ such that any two words
$w,w'\in\Sigma\scc{-t^\to_f}{t^\to_f}$ with the same trace $T^{\cc{-1}0}_f(w)=T^{\cc{-1}0}_f(w')$ have the same letter $w_1=w'_1$ just on the right.

If we iterate this with a growing trace size, we can rebuild all letters of the right half of the initial word: for any $n\in\N$ and any $y\in T_f^{\cc{-1}0}(\Sigma\scc{-n}n)$, there exists a unique $v\in\Sigma\scc1{\ipart{n/t^\to_f}}$ such that any $w\in\Sigma\scc{-n}n$ with $T^{\cc{-1}0}_f(w)=y$ satisfies $w\scc1{\ipart{n/t^\to_f}}=v$.
This bijection gives in particular that the entropy of an expansive CA $f$ (with $\Sigma\scc{-k}k=A^{\cc{-k}k}$) is at least $\frac{\log\card A}{t^\to_f}$.

Intuitively, if we consider some computation triangle where both the right part $w\scc0n$ of the initial configuration and the trace $T_f(w)$ are fixed, then it is clear that there is always at most one way to complete the right part of the triangle. In the right-expansive case, there is also at most one way to complete a portion $w\scc{-\ipart{n/{t^\to_f}}}0$ of the left part.
 
We say that the subautomaton of some CA is \emph{right-permutive} if it is the subautomaton of some (possibly different) right-permutive CA. This implies that it is right-expansive.
Symmetrically, we can define \emph{left expansive} CA or subautomata, with some particular time step $t^\leftarrow_f$, and \emph{left-permutive} CA or subautomata with $t^\leftarrow_f=1$. A CA or subautomaton is \emph{expansive} if it is both left and right expansive.
It is \emph{bipermutive} if it is both left-permutive and right-permutive.

Now the definition of $t^\to_f$ helps us build large fooling sets.
\begin{lem}\label{l:exp0}
 Let $f\restr\Sigma$ be an expansive subautomaton of some CA, $z\in A^{n+1}$, and
 \[W_z=\set w{\Sigma\scc{-\ipart{\frac n{t^\leftarrow_f}}}{\ipart{\frac n{t^\to_f}}}}{\exists x,y,xwy\in\Sigma\scc{-n}n,T_f(xwy)=z}~.\]
 Then the multi-round CC of $\hat f_z$ is lower-bounded by $\log\card{W_z}$.
\end{lem}
\begin{pf*}{Proof.}
For any $w\in W_z$, let us define $\gamma(w)=(x_ww\scc{-\ipart{n/t^\leftarrow_f}}{-1},w\scc1{\ipart{n/t^\to_f}}y_w)$, where $x_w$ and $y_w$ are fixed words such that $T_f(x_wwy_w)=z$.
Note that $\gamma$ is injective, with $\gamma^{-1}(u,v)=u\scc{-\ipart{n/t^\leftarrow_f}}{-1}z_0v\scc1{\ipart{n/t^\to_f}}$.
Moreover, let us show that $\gamma(W_z)$ is a fooling set for $\hat f_z$.
By construction, if $w\in W_z$, then $\hat f_z(\gamma(w))=1$.
Now let $w'\in W_z$ such that $T_f(x_ww\scc{-\ipart{n/t^\leftarrow_f}}0w'\scc1{\ipart{n/t^\to_f}}y_{w'})=z=T_f(x_wwy_w)$. Right expensiveness will give that the initial configurations $w\scc1{\ipart{n/t^\to_f}}y_w$ and $w'\scc1{\ipart{n/t^\to_f}}y_{w'}$ of the two triangles begin equally: $w\scc1{\ipart{n/t^\to_f}}=w'\scc1{\ipart{n/t^\to_f}}$. If besides $T_f(x_{w'}w'\scc{-\ipart{n/t^\leftarrow_f}}{-1}w\scc0{\ipart{n/t^\to_f}}y_w)=z$, then symmetrically, left expensiveness gives that $w\scc{-\ipart{n/t^\leftarrow_f}}{-1}=w'\scc{-\ipart{n/t^\leftarrow_f}}{-1}$. We globally obtain that $w=w'$, hence $\gamma(w)=\gamma(w')$, \ie $\gamma(W_z)$ is a fooling set.
Thanks to Proposition \ref{p:fooling}, the CC is at least $\log\card{\gamma(W_z)}=\log\card{W_z}$.
\qed\end{pf*}
Our interest will be that when $W_z$ is sufficiently large, the CC is linear.
If we study combinatorially the set of all possible traces, we will see conditions for it to be large.
\begin{lem}\label{l:exp1}
 Let $f\restr\Sigma$ be an expansive subautomaton of some CA and $k\in\co1{\card{\Sigma\scc1n}}$.
If $p$ is the number of words $z\in A^{n+1}$ such that the multi-round CC of $\hat f_z$ is more than $\log k$, then:
\[p\ge\frac{\card{\Sigma\scc{-\ipart{n/t^\leftarrow_f}}{\ipart{n/t^\to_f}}}-k}{\card{\Sigma\scc1n}-k}~.\]
\end{lem}
\begin{pf*}{Proof.}
For $z\in A^{n+1}$, consider $W_z$ as defined in Lemma \ref{l:exp0}, and for $w\in W_z$, $\pi(w)=w\scc1n$. Note that if $\pi(w)=\pi(w')$, since $T_f(w)=T_f(w')=z$ and $f$ is left-expansive, then we have $w=w'$. It results that $\card{W_z}=\card{\pi(W_z)}\le\card{\Sigma\scc1n}$.
Moreover, consider the number $q$ of words $z\in A^{n+1}$ such that $W_z$ admits more than $k$ elements. Then we can distinguish between the sets $W_z$ these $q$ bigger ones (which have cardinality at most $\card{\Sigma\scc1n}$ as stated above), with the other, smaller, ones (which have cardinality at most $k$):
\[\sum_{z\in A^{n+1}}\card{W_z}=\sum_{\card{W_z}>k}\card{W_z}+\sum_{\card{W_z}\le k}\card{W_z}\le\sum_{\card{W_z}>k}\card{\Sigma\scc1n}+\sum_{\card{W_z}\le k}k~.\]
We obtain:
\[\sum_{z\in A^{n+1}}\card{W_z}\le q\card{\Sigma\scc1n}+(\card A^{n+1}-q)k~.\]
On the other hand, we have $\bigcup_{z\in A^{n+1}}W_z=\Sigma\scc{-\ipart{n/t^\leftarrow_f}}{\ipart{n/t^\to_f}}$ (since every word has a trace), hence: \[\sum_{z\in A^{n+1}}\card{W_z}\ge\card{\Sigma\scc{-\ipart{n/t^\leftarrow_f}}{\ipart{n/t^\to_f}}}~.\]
Putting the two inequalities together, we get:
\[q\card{\Sigma\scc1n}+(\card A^{n+1}-q)k\ge\card{\Sigma\scc{-\ipart{n/t^\leftarrow_f}}{\ipart{n/t^\to_f}}}~.\]
As a result, \[q\ge\frac{\card{\Sigma\scc{-\ipart{n/t^\leftarrow_f}}{\ipart{n/t^\to_f}}}-\card A^{n+1}k}{\card{\Sigma\scc1n}-k}~.\]
By Lemma \ref{l:exp0}, for any of the $q$ words with $\card{W_z}\ge k$, the multi-round CC of $\hat f_z$ is at least $\log k$.
\qed\end{pf*}
By symmetry, $\Sigma\scc{-n}{-1}$ may replace $\Sigma\scc1n$ in the previous formula.

Let us first see the case of a CA (without forbidden patterns).
\begin{prop}\label{p:exp}
 If $f$ is an expansive CA with $m=1/t^\to_f+1/t^\leftarrow_f-1>0$ and $n>0$, then there exists some word $z\in A^{n+1}$ such that the multi-round CC of $\hat f_z$ is lower-bounded by $nm\log\card A$.
\end{prop}
\begin{pf*}{Proof.}
We just use Lemma \ref{l:exp1} with $\card\Sigma\scc{-\ipart{n/t^\leftarrow_f}}{\ipart{n/t^\to_f}}=\card A^{nm+n+1}$, $\card{\Sigma\scc1n}=\card{A^{\cc1n}}=\card A^n$ and $k=1$.
We obtain that the number $p$ of words $z\in A^{n+1}$ such that the multi-round CC of $\hat f_z$ is more than $0=\log 1$ is $p\ge\card A^{n+1}\frac{\card A^{nm}-1}{\card A^n-1}>0$.
% For $z\in A^{n+1}$, let us define $W_z$ as in Lemma \ref{l:exp0}. %=\sett{w\scc{-n/t^\leftarrow_f}{n/t^\to_f}}{T_f(w)=z}$.
% We have $\bigcup_{z\in A^{n+1}}W_z=A^{\cc{-n/t^\leftarrow_f}{n/t^\to_f}}$ (since every word has a trace), hence $\sum_{z\in A^{n+1}}\card{W_z}\ge\card A^{nm+n+1}$.
% The average cardinal $\card{W_z}$ over all possible words $z$ is thus at least $\card A^{nm}$.
% There is at least one $z$ for which $\card{W_z}$ is at least this average value.
% We conclude by Lemma \ref{l:exp0}.
\qed\end{pf*}
The previous result cannot hold for all possible words $z\in A^{n+1}$, since there are expansive (not bipermutive) CA for which some of these words do not appear in $T_f(A^{\cc{-n}n})$, and hence correspond to a trivial CC.
In other words, in the case of a large expensiveness speed on both sides, the CC is linear for some words; if we allow an arbitrarily low linearity constant, Lemma \ref{l:exp1} can actually give rather large families: if $0\le s<m$, then the multi-round CC of $\hat f_z$ is more than $ns\log\card A$ for at least $\card A^{n+1}\frac{\card A^{nm}-\card A^{ns}}{\card A^n-\card A^{ns}}$ words.
The same inequalities hold when the CA is permutive on one side and expansive on the other one.
In the particular case of bipermutivity, we have a linear traced CC associated to any word.
\begin{prop}
 For any bipermutive CA and any word $z\in A^{n+1}$, the multi-round CC of $\hat f_z$ is equal to $n\log A$.
\end{prop}
\begin{pf*}{Proof.}
%  To apply Lemma \ref{l:biperm}, it is enough to take $U=V=A^n$.
Just apply Lemma \ref{l:exp1} with $\Sigma$ full as in the previous proof, $t^\to_f=t^\leftarrow_f=1$ (\ie $m=1$), and $k=\card A^n-1$.
We get that the number $p$ of words $z\in A^{n+1}$ such that the multi-round CC of $\hat f_z$ is more than $\log k$ is $p\ge\card A^{n+1}\frac{\card A^{nm}-k}{\card A^n-k}=\card A^{n+1}$, \ie all words of $A^{n+1}$ have a CC of at least $n\log{\card A}$.
The converse inequality is obvious.
\qed\end{pf*}
The expansive elementary CA are exactly the four bipermutive ones (90, 150, 105, 165) and have thus the maximal possible traced CC for all words.
%For other expansive CA, it will be linear but with a constant depending on $t^\to_f$ and $t^\leftarrow_f$. These two characteristic times do not experimentally seem ever big -- hence it gives a rather good lower bound for the CC -- but are not theoretically very well understood: their computability with the CA as input is still an important open question in CA theory (see \cite{jarkko}).
\fig{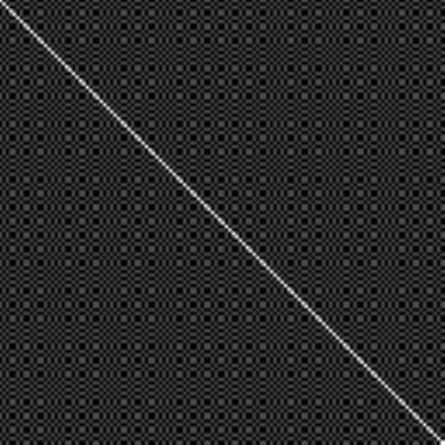}{Matrix of the rule 90}{.5}

Proposition \ref{p:exp} involves only CA which have an expensiveness speed of more than a half; we will now see that it actually represents the best limit of expensiveness speed we could get for this result.

It is not difficult to observe that the $2$-grouped of some expansive CA $f$ is still expansive, with $t^\leftarrow_{f\bk2}=2t^\leftarrow_f$ and $t^\to_{f\bk2}=2t^\to_f$. Hence we have the following example of expansive CA which is simple with respect to traced CC.
\begin{exmp}
 Consider the CA on alphabet $\cc03$ defined by the local rule (where $/$ is the quotient of the Euclidean division):
 \[\appl f{\cc03^3}{\cc03}{(a,b,c)}{2((a+b)\bmod2)+((b/2+c/2)\bmod 2)~.}\]
 If we identify $\cc03$ with $\{0,1\}^2$, this CA is the $2$-grouped of the bipermutive CA $90$.
 Hence it has $t^\leftarrow_f=t^\to_f=1/2$. On the other hand the traced CC of any word is $1$, by Proposition \ref{p:bulk} (the converse inequality is rather obvious).
\end{exmp}

\subsection{Legal rules}
We now see another little application of Lemma \ref{l:exp0}, in the case of binary alphabet.

For $\cc ij\subset\Z$ and $u\in A^{\cc ij}$, let us denote $\rev u\in A^{\cc{-j}{-i}}$ the \emph{mirror} of $u$, \ie the word such that $\rev u_{-k}=u_k$ for any $k\in\cc ij$.
If $V\subset A^{\cc ij}$, we note $\rev V\subset A^{\cc{-j}{-i}}$ the set of all mirrors of words of $V$.
The subautomaton $f\restr\Sigma$ of some CA is \emph{$0$-legal}, with $0\in A$, if for any $u\in\Sigma\scc{-1}1$, $\rev u\in\Sigma\scc{-1}1$ and $f(\rev u)=\rev{f(u)}$, and for any $a\in A$ such that $a0a\in\Sigma\scc{-1}1$, $f(a0a)=0$.

\begin{lem}\label{l:symmetric}
 For any $0$-legal subautomaton $f\restr\Sigma$ of some CA, any $u$ such that $\rev u0u\in\Sigma\scc{-n}n$, and any $t\le n$, $f^t(\rev u0u)_0=0$.
\end{lem}
\begin{pf*}{Proof.}
 This comes from an immediate recurrence: if $n>0$, then $f(\rev u0u)=f(\rev u0)f(u_00u_0)f(0u)=\rev{f(0u)}0f(0u)$, which still has the same form.
\qed\end{pf*}

The previous lemma allows in this context to establish an equivalence between the problem of the traced CC and the classical equality test of binary words, which is known to have linear CC.
\begin{prop}\label{p:permsym}
 Let $f\restr\Sigma$ be a $0$-legal bipermutive subautomaton of some CA $f$ such that $\Sigma\scc{-n}n$ contains some sublanguage of the form $\rev V0V$.
 Then the multi-round CC of $\hat f_{0^{n+1}}$ is at least $\log\card V$.
\end{prop}
  \begin{pf*}{Proof.}
Just apply Lemma \ref{l:exp0}, with $W_{0^{n+1}}\supset\rev V0V$ thanks to Lemma \ref{l:symmetric}. We get a traced CC of at least $\log\card{W_{0^{n+1}}}=\log\card{\rev V0V}=\log\card V$.
\qed\end{pf*}

\begin{cor}
 The CA 18, 26, 146, 154, 218 have a multi-round CC in $\Omega(n)$. %26=82, 154=210
\end{cor}
\begin{pf*}{Proof.}
Note that these rules are equal to the bipermutive rule 90 except on neighborhoods $011,110,111$.
Define $\Sigma$ as the set of words avoiding the pattern $11$ and the patterns $10^{2k}1$, for $k\in\N$.
It can be easily seen that $\Sigma$ is stable by the synchronous application of CA 90, hence by any of these CA.\\
Let $m\in\N$ and $V=(0100+0001)^m$ (standard notation for languages, seen as words indexed in $A^{\cc1{4m}}$). Note that $\rev V0V\subset\Sigma\scc{-4m}{4m}$.
 From Proposition \ref{p:permsym}, the multi-round traced CC corresponding to $f\restr\Sigma$ and $z=0^{4m+1}$ is greater than $\log\card V=m$.
\qed\end{pf*}
\fig{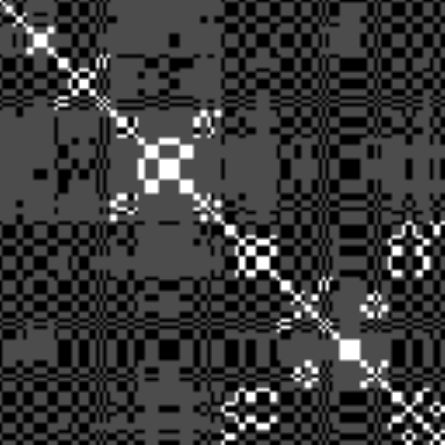}{Matrix of the rule 146}{.5}
Unfortunately, we do not know similar practical subsystems for the three other bipermutive elementary CA. %that would allow other applications of Proposition \ref{p:permsym}.
%150 aussi a un sous-système: (001((00)*)100+(00)*+(11)*)*
%165: (01(11)*)*

% \section{Simulation}
% 
% By simulation, we mean a composition of temporal grouping (of several time steps together), of spacial grouping (of several cells together), and of subautomata operations. There exist so-called \emph{intrinsically universal} CA, which simulate any other CA.
% It is known from \cite{ccuniv} that the CC of the predication, invasion and cycle problems are decreasing with respect to simulation.
% For subautomata, what makes things work is the fact that all letters are involved in the definition of the problem.
% 
% For our purpose, we could very well imagine an intrinsically universal CA for which the problem $\hat f_z$ is easy for some word $z$ (for instance for $0^{n+1}$, where $0$ is a fresh letter artificially added to some already intrinsically universal CA).
% Nevertheless, the notion can be made decreasing with respect to simulation if we take into account any subalphabet.
% For instance, let us take as a complexity measure the maximal CC among all problems $\hat f_{a^{n+1}}$ for $a\in A$. Then any simulated CA has a lower complexity, up to some multiplicative constant. In particular, the complexity for such intrinsically universal CA must be in $\Omega(n)$. This observation might be helpful to prove that some specific CA are not intrinsically universal, as in \cite{ccuniv}.

\section*{Conclusion}

In this paper, we have addressed a new problem of communication complexity to get some clues about the information streams present in the evolution of CA. This can help understand their behaviors by exhibiting how much communication is needed to achieve their computation.

We have treated a large number of elementary cellular automata; some of the remaining ones look experimentally simple, other ones rather mysterious, such as 22, which nearly has a bipermutive subautomaton, or as some of the CA for which $1$ is weakly but not semi-strongly spreading.

Unlike the classical CC of CA, this notion of complexity does not a priori present links with cellular simulation (see Proposition \ref{p:bulk}). This could be overpassed by defining a more general problem, where Alice and Bob would need to determine whether the trace belongs to some given subset of $A^{n+1}$ or not. This extends both classical (at least in the binary case) and traced CC, and one should carefully consider what kind of subsets would imply a good notion of complexity for CA.

On the contrary, our approach allows more links with topological dynamics than classical CC. In \cite{KurkCA}, Petr K\r urka classified the CA into four classes: equicontinuous, almost equicontinuous (and not equicontinuous), sensitive (and not expansive) and expansive. We have proved that the first one implies a trivial CC and a strong version of the last one a very complex one. In the construction of the fooling sets of Section \ref{s:posexp}, the ability to reconstruct the initial word is crucial; maybe if we ask $\hat f_z$ to be complex for any word $z$, it would imply something close to expensiveness (our condition being then nearly necessary). When fixing the word $z$, a high complexity for the problem is not possible without a large set of initial words on the right and on the left, independent from each other, and which together can give $z$ in the trace.

K\r urka's intermediary classes do not imply anything on this kind of complexity. Nevertheless, almost equicontinuity may be related to a simple \emph{average CC}, since in that case ergodic theorists know that almost the whole system behaves as an equicontinuous system. Understanding this distinct complexity measure could be a track for future research.

% The following elementary CA are not concerned by any of the previous results:
% \begin{itemize}
% \item 23, 151: alternating $0$s and $1$s
% \item 5, 219, 1, 201: equicontinuous
% \item 91
% \item 37, 133
% \item 22: linear?
% \item 9 (65), 73, 41 (97), 129, 137 (193), 161, 169 (225), 233
% \item 33: logarithmic?
% \end{itemize}

% Another perspective would be to widen the radius. This would introduce the problem of considering a single cell, but thanks to synchronism, we can hope that the evolution of each cell cannot be much easier to compute than the evolution of a group of cells.

\section*{Acknowledgement}
Thanks to the referees for their careful reading and precise correction, from a rather rough version of that work.

\bibliographystyle{entcs}
\bibliography{cc}

\end{document}